\documentclass[10pt,twocolumn,preprintnumbers,superscriptaddress,nofootinbib,aps,prd]{revtex4-2}

\usepackage{graphicx}
\usepackage{amsmath}
\usepackage{srcltx}
\usepackage[utf8]{inputenc}
\usepackage[colorlinks]{hyperref}
\usepackage{times}
\usepackage{amssymb}
\usepackage{slashed}

\newcommand{\lsim}{\lesssim}
\newcommand{\gsim}{\gtrsim}

\newcommand{\eq}[1]{Eq.~(\ref{#1})}

\newcommand{\ord}[1]{\mathcal{O}{(#1)}}
\newcommand{\beq}{\begin{equation}}
\newcommand{\eeq}{\end{equation}}
\newcommand{\bea}{\begin{eqnarray}}
\newcommand{\eea}{\end{eqnarray}}

\newcommand{\mP}{M_{\rm P}}

\newcommand{\tmax}{T_{\rm max}}

\newcommand{\appropto}{\mathrel{\vcenter{
  \offinterlineskip\halign{\hfil$##$\cr
    \propto\cr\noalign{\kern2pt}\sim\cr\noalign{\kern-2pt}}}}}

\begin{document}

% page numbers bottom-center
\pagestyle{plain}

\title{\boldmath Adagio for Thermal Relics}

\author{Hooman Davoudiasl}
\email{hooman@bnl.gov}

%\author{Mathew Sullivan}
%\email{}

\author{Matthew Sullivan}
\email{msullivan1@bnl.gov}

\affiliation{High Energy Theory Group, Physics Department \\ Brookhaven National Laboratory, Upton, NY 11973, USA}

%%%%%%%%%%%%%%%%%%%%%%%%%%%%%%%%%%%%%%%%%%%%%%%%%%%%%%%%%%%%%%%%%%%%%%%%%%%%

\begin{abstract}
	
A larger Planck scale during an early epoch leads to a smaller Hubble rate, which is the measure for efficiency of primordial processes.  The resulting slower cosmic tempo can accommodate alternative cosmological histories.  We consider this possibility in the context of extra dimensional theories,  which can provide a natural setting for the  scenario.  If the fundamental scale of the theory is not too far above the weak scale, to alleviate the ``hierarchy problem," cosmological constraints imply that thermal relic dark matter would be at the GeV scale, which may be disfavored by cosmic microwave background measurements.  Such dark matter becomes viable again in our proposal, due to smaller requisite annihilation cross section, further motivating ongoing low energy accelerator-based searches. Quantum gravity signatures associated with the extra dimensional setting can be probed at high energy colliders -- up to $\sim 13$ TeV at the LHC or $\sim 100$ TeV at FCC-hh. Searches for missing energy signals of dark sector states, with masses $\gsim 10$ GeV, can be pursued at a future circular lepton collider.

\end{abstract}
\maketitle

%%%%%%%%%%%%%%%%%%%%%%%%%%%%%%%%%%%%%%%%%%%%%
%%%%%%%%%%%%%%%%%%%%%%%%%%%%%%%%%%%%%%%%%%%%%

\section{Introduction}

Cosmological observations of light element abundances have led us to conclude that our understanding of the cosmos, based on the Standard Model (SM) and general relativity, can provide a quantitative description of the Universe when it was a few seconds old \cite{ParticleDataGroup:2022pth}.  This corresponds to the era of Big Bang Nucleosynthesis (BBN), dominated by radiation at a temperatures of $\ord{\rm MeV}$. The agreement between theory and observation implies that the rate of the cosmic expansion in this era, given by the Hubble parameter $H$, is set by a plasma that cannot have significant contribution from unknown physics.       

At the same time, the study of cosmology has provided us with some of the starkest clues that our fundamental understanding of the Universe remains incomplete.  Key examples are the mystery of what constitutes dark matter (DM) and how visible matter evaded complete annihilation, {\it i.e.} what is the source of the cosmic baryon asymmetry.  There is broad consensus in particle physics and cosmology that new fundamental ingredients are needed to address these questions.  While a great variety of ideas have been proposed to explain either problem, none has been shown to be a definitive resolution, neither empirically  nor via inescapable theoretical imperatives.

The quantitative understanding of the cosmos back to the BBN era illustrates that the predicted rate for the relevant microscopic processes, when compared to the expansion rate of the Universe at the corresponding epoch, are generally correct within margins of error.  Similarly, the efficiency of a new physics mechanism that aims to address a cosmological puzzle is  measured against the expansion rate set by $H$.  In general relativity, the expansion rate itself is set by the gravitational response of spacetime to various forms of energy density and (assuming zero curvature) $H \propto \mP^{-1}$, where $\mP \approx 1.2 \times 10^{19}$~GeV \cite{ParticleDataGroup:2022pth} is the Planck mass set by Newton's constant $G_N=\mP^{-2}$ (with the reduced Planck constant $\hbar$ and the speed of light $c$ set to unity: $\hbar=c=1$). 

The above account implies that if gravity had a different strength at very early times, expectations about the viability of a cosmological model would change.  In particular, if gravity was much weaker well before the BBN era, certain processes that are deemed too inefficient may have been sufficiently fast.  A good example is provided by thermal relic DM with parameters that lead to inefficient annihilation, resulting in its  overabundance and conflict with precision cosmological data.  One may address this problem in various ways, for example by generating additional entropy later on to dilute the DM density \cite{Davoudiasl:2015vba}.  However, we will entertain a less explored possibility here, namely {\it weaker gravity leading to a slower expansion}, which allows a longer time for DM to annihilate before its abundance is set by freeze-out.  To realize this `Adagio' scenario, we will assume that during some early cosmological epoch, the value of the Planck mass was larger, reducing the strength of gravitational coupling.

The key feature of our scenario is the time variation of gravitational coupling in the early stages of cosmology, which makes the Planck mass a function of time: $\mP = \mP(t)$.  For  $t<t_*$, corresponding to temperatures $T > T_*$, gravity was weaker, {\it i.e.} $\mP(t) > \mP(t_*)$, and we will assume that $\mP$ was constant afterwards.  In general, we would like $t_*$ to be early enough that the well-established features of the early Universe are not perturbed significantly.  As we will illustrate, this can be achieved as long as $t_*$ is sufficiently small compared to $\sim1$~s ($T_*$ is sufficiently large compared to $\sim 1$~MeV) corresponding to the onset of BBN, which we will assume to go through according to the standard theory. 

One may simply postulate that $\mP$ had some temperature dependence, $\mP\to \mP(T)$, that led to its variation over cosmic time, and the value that we observe today was set before the BBN epoch.  This is not the way we usually think of gravity, but in fact such a behavior for $\mP$ could be realized in theories with $n\geq 1$ extra dimensions (an early suggestion for this possibility can be  found in Ref.~\cite{Kamionkowski:1990ni}).  In an extra-dimensional framework -- which is generally deemed  necessary for a proper formulation of quantum gravity in string theory -- the observed 4-d $\mP$ is a derived quantity.  The true fundamental scale $M_F$ in $(4+n)$-d may be much smaller than  $\mP$ if the extra dimensions are compact, with a typical size $R \gg M_F^{-1}$ \cite{Arkani-Hamed:1998jmv,Antoniadis:1998ig}, according to 
\beq
\mP^2\sim M_F^{2+n} R^n\,.
\label{ADD}
\eeq
 
On general grounds, one may expect that the extra dimensions are initially small, {\it i.e.} $\sim M_F^{-1}$, but dynamically grow to become large.  This would then translate to a time variation of 4-d gravitational coupling set by $\mP$.  Given the above, we will adopt $(4+n)$-d theories as our basic underlying framework.  Such models can in principle alleviate  the ``hierarchy problem" -- {\it i.e.}, the smallness of the Higgs mass $m_H \approx 125$~GeV compared to the implied scale of gravity  -- by lowering $M_F$ to be not far above $m_H$.

As we will discuss in the following, the relative proximity of $M_F$ to the TeV scale generally constrains the maximum reheat temperature of the Universe to $\lsim$~GeV, which would point to masses for thermal relic DM at the GeV scale.  This regime of masses has garnered significant attention in recent years, as an alternative to the traditional weak scale models, characterizing dark sectors that are potentially accessible to a wide range of laboratory experiments (see, for example, Ref.~\cite{Krnjaic:2022ozp}).  Note that large classes of thermal relic models of DM with a mass $\lsim 10$~GeV are ruled out by Cosmic Microwave Background (CMB) observations \cite{Madhavacheril:2013cna,Planck:2018vyg}, making them less motivated as experimental targets.   However, if DM annihilation cross sections can be lowered significantly, as in our scenario, those models can become viable again.

Next, we will describe the main features of our $(4+n)$-d framework and sketch how the early Universe evolution, leading to variable gravity, unfolds in our scenario.  We will then consider the implications of our model for DM production and outline some of its phenomenological consequences.  A summary and some concluding remarks are given at the end of this work.                       

%%%%%%%%%%%%%%%%%%%
\section{Models with Extra Dimensions}
%%%%%%%%%%%%%%%%%%%

In principle, the scale of compactification of $n$ additional dimensions could be very high, which would make the underlying physics inaccessible to low energy measurements. 
One could assume that the fundamental scale of the $(4+n)$-d theory is not very far from the weak scale $\sim \ord{\rm TeV}$, potentially addressing its origin.  This will be the main scenario we will consider in what follows, for it can be motivated as a resolution of the hierarchy problem and may be probed in future high energy experiments.

We will adopt the general picture described in Ref.~\cite{Arkani-Hamed:1999fet} as the starting point of the cosmological evolution.  Initially, all spatial dimensions have a size $\gsim M_F^{-1}$.   The basic idea is that one can construct a model of inflation that satisfies key  observational  requirements, where the initial inflationary era leads to rapid expansion along the visible {\it 3-brane} dimensions, while the compact dimensions remain fixed near the fundamental size $\gsim M_F^{-1}$.  After  this main inflationary era ends, the size of the extra dimensions, governed by the radion potential, starts to grow.  During this time, the non-compact dimensions  shrink and the radiation contained on the visible brane blue-shifts.  Once the radiation on the brane and radion potential have equal sizes, the contraction of the visible dimensions would stop.  

In typical scenarios for which Ref.~\cite{Arkani-Hamed:1999fet} aims to provide a cosmological framework, the radion eventually reaches the minimum of its potential where the extra dimensions attain their final stabilized size.  After this point, the evolution of Universe will resemble the standard 4-d picture, at low energies.  A generic problem in this scenario is that the radion ends up as an oscillating modulus which is long-lived on cosmological time scales and can lead to early matter domination and possible conflict with cosmological data.  This could be addressed, for example, by a brief period of secondary inflation, diluting the energy density in the radion field \cite{Arkani-Hamed:1999fet}. 

We will consider  the framework of Ref.\cite{Arkani-Hamed:1999fet}, outlined above, but depart from it by adding an interlude to the evolution of the cosmos before it ends up with stable extra dimensions.  In our modified scenario, the radion potential initially has a different minimum corresponding to larger compact dimensions than arrived at eventually.  This  intervening cosmological era would then be governed by  a 4-d gravitational interaction that could have been significantly weaker than the one observed today.  Below, we will argue that the general demands of our proposed scenario do not fit within the specific model assumed in Ref.~\cite{Arkani-Hamed:1999fet}, where the radion potential acts as the main source of inflation.

Assuming that the radion potential $V_{\rm rad}(R)$ is the source of inflation along the 3-brane dimensions,  corresponding to the visible world, the primordial density perturbations are given by \cite{Arkani-Hamed:1999fet} 
\beq
\frac{\delta \rho}{\rho} \approx 
\frac{5}{12 \pi\, (M_F R_I)^n S}\left[\frac{V_{\rm rad}(R_I)}{12 n(n-1)\, M_F^4}\right]^{1/2}\,,
\label{delrho}
\eeq
where $R_I$ is the size of the extra dimensions during inflation and $S$ is a parameter that needs to be $\ord{10^{-3}}$ for a consistent inflationary scenario.  
To avoid significant deviations from scale-invariant perturbations, as required by data, $R_I$ should be approximately constant, $R_I\sim M_F^{-1}$, during inflation.  

As we will discuss below, the largest reheat   temperature $\sim \ord{\rm GeV}$, consistent with cosmological constraints, can be attained for maximal $n$, and so we will mostly focus on the case with $n=6$ extra dimensions and $M_F \gsim 10$ TeV, for general consistency with experimental bounds that we will discuss later.  This implies that in order to have a suppressed  Hubble constant during freeze-out, the radion potential may only transition to its ``late" Universe minimum, corresponding to the present value of $\mP$, at $T<$~GeV.  This would typically demand that $V_{\rm rad}$ is governed by scales $\lsim \ord{\rm GeV}$.  Using \eq{delrho}, the preceding  considerations imply that $\delta \rho/\rho \lsim 7\times 10^{-8}$, which is well below the measured value $\sim \ord{10^{-5}}$ \cite{ParticleDataGroup:2022pth}.  Note that here, the spectral index $n_s$ is given by \cite{Arkani-Hamed:1999fet}  
\beq
1-n_s\lsim n(n+2) \ord{S}.
\label{1-ns}
\eeq  
Since the measured value is $n_s \approx 0.97$ \cite{Planck:2018vyg}, choosing $S\ll 10^{-3}$ is not a viable option for enhancing the density perturbations in \eq{delrho}.

Given the above analysis, we then assume that an appropriate brane-localized potential is present for an inflaton $\Phi$, such that it allows for sufficient inflation of $\sim 60$ e-folds, or more, to address large scale features of the cosmos.  The density perturbations produced during the slow roll of $\Phi$ are given by 
\beq
\frac{\delta \rho}{\rho} \sim 
\frac{H_I^2}{\dot \Phi}\,,
\label{delrhoII}
\eeq
where 
\beq
H_I^2 \sim \frac{V(\Phi)}{M_F^{n+2} R_I^n}
\label{HI2}
\eeq
gives the inflationary Hubble scale and  $\dot \Phi$ is the time derivative of $\Phi$.  During inflation, $3 H_I \dot \Phi\approx -dV(\Phi)/d\Phi$, which is subject to the slow-roll condition 
\beq
\frac{dV(\Phi)}{d\Phi} \lsim \frac{V(\Phi)} {M_F (M_F R_I)^{n/2}}.
\label{slowroll}
\eeq
We thus have 
\beq
\frac{\delta \rho}{\rho} \gsim 
\frac{V(\Phi)^{1/2}}{M_F^2 (M_F R_I)^n}.
\label{delrhoIII}
\eeq
For $M_F R_I \sim 1$, choosing $V(\Phi)^{1/2} \sim (100~{\rm GeV})^2$ can then easily yield the observed level of density perturbations.

Based on the above discussion, we may then assume that the radion potential stays at a minimum that yields $R > R_0$, where $R_0$ is today's size of extra dimensions, until after freeze-out at $T<$~GeV.  We note that for
\beq
\kappa \equiv  \frac{R_{\rm max}}{R_0}\,, 
\label{kappa}
\eeq
the intermediate value of Planck mass would be $\kappa^{n/2}$ times larger and the corresponding Hubble constant would be smaller by that amount.  This would allow consideration of thermal relic DM with $\ord{10}$ times smaller annihilation cross sections than in the standard picture, for $\kappa \sim 2$ and $n=6$.

\subsubsection{Constraints on Extra Dimensional Cosmology}

During the period where the compact extra dimensions are changing size, the $(4+n)$-dimensional metric is approximately described by the Kasner solutions \cite{Arkani-Hamed:1999fet},
\beq
    \label{kasner}
    ds^2 = dt^2 - a_i^2 \left(\frac{t}{t_i}\right)^{2k} d\vec{x}_3^2 - b_i^2 \left(\frac{t}{t_i}\right)^{2l} d\vec{y}_n^2 ,
\eeq
where $a_i$ and $b_i$ are the initial scale factors for the 3 large and $n$ compact dimensions, respectively, and $t_i$ is the initial time where the contraction of the compact extra dimensions begins. For the case where the compact dimensions are contracting, the values of $k$ and $l$ in the exponents are given by
\begin{eqnarray}
k &=& \frac{3+\sqrt{3n(n+2)}}{3(n+3)} \nonumber \\
l &=& \frac{n-\sqrt{3n(n+2)}}{n(n+3)}.
\end{eqnarray}
Note that these are the solutions with opposite sign from those considered in Ref.~\cite{Arkani-Hamed:1999fet}. For $n=6$ extra dimensions, we obtain $k=5/9$ and $l=-1/9$. This implies that if the compact dimensions shrink by a factor of $\kappa$, then the large dimensions will increase in size by a factor of $\kappa^5$. Since the temperature cools as the 3-dimensional universe expands, we require that the anomalous expansion from the Kasner phase ends before the temperature  falls below $T_{\rm BBN}\approx 2$~MeV ~\cite{Hannestad:2004px,Hasegawa:2019jsa}, to avoid significant deviation from standard BBN. This means the Kasner phase must begin before 
\beq
\label{Tmin}
T_{\rm min} \approx \kappa^{|k/l|} \,T_{\rm BBN}.
\eeq

The presence of large extra dimensions allows for production of light Kaluza Klein (KK) gravitons in the early Universe which could cause conflict with observational data.  To avoid such problems, one is led to assume that the Universe did not attain a high reheat temperature, which limits the scope of cosmological models considered in this framework. 
 These considerations were revisited in Ref.~\cite{Macesanu:2004gf}, and the most stringent constraint, based on preserving the products of the BBN, was determined to be   
\beq
\left(\frac{r \, T_{\rm max}}{M_F}\right)^{n+2}\lsim
10^{-31},
\label{Tmax-MT}
\eeq
where $r\approx 6$ is a numerical factor, and $\tmax$ is the maximum reheat temperature.  

To adapt the above bound (\ref{Tmax-MT}) to our Adagio scenario, we multiply the left-hand side by a factor $\kappa^{n/2}$ to account for the fact that we assume a value of $\mP\propto R^{n/2}$ during the relevant cosmological era that is $\kappa^{n/2}$ times larger (to avoid excessively complicated results, we have only considered this factor that gives the main effect for general $n$). An  additional factor of $\kappa$ should also be included to reflect the growth of the graviton KK mass scale by $\sim \kappa$ after the extra dimensions shrink to their late Universe size; the more massive the relic that decays, the more stringent the bound from BBN on its abundance \cite{Macesanu:2004gf}.  With these  modifications, we obtain the following relation that applies to our scenario
\beq
\tmax \lsim \frac{10^{-31/(n+2)}}{r\,\kappa^{1/2}}\,M_F. 
\label{Tmax}
\eeq
 The above bound could  be somewhat alleviated if one accounts for the   non-hadronic decay channels of KK gravitons    \cite{Macesanu:2004gf}, but we adopt it to be more conservative.   As can be seen from  \eq{Tmax}, the dependence of $\tmax$ on $\kappa\lsim 10$ is not very strong. Requiring $T_{\rm min}<T_{\rm max}$ leads to
\beq
M_F \gsim r\, \kappa^{|k/l|+1/2} \,10^{31/(n+2)}\, T_{\rm BBN}.
\eeq
 
The temperature where the radion potential readjusts to its  late value will be taken to be well below the freeze-out temperature which implies $T_* \ll \tmax$.  This allows for a simpler and more transparent treatment of the cosmic evolution in our work.  Hence, we can  assume that the DM relic abundance is set while the value of $\mP$ is larger than today's value,  but constant.

\section{Demonstration with Dark Photon Mediated Dark Matter}

For the purpose of demonstration, we present an analysis using a concrete dark matter model with fermionic dark matter coupled to a dark photon, associated with a dark  $U(1)_D$ gauge interaction, which kinetically mixes with the SM photon.  We will assume that the dark sector is localized on the same brane as the SM content, making it effectively 4-dimensional.  For more general treatments of dark photon and kinetic mixing in extra dimensional models see, for example, Refs.~\cite{Rizzo:2018ntg,Rizzo:2018joy,Rizzo:2020ybl}, where other possible effects may also allow circumvention of the CMB bounds considered here.

The phenomenologically relevant part of the Lagrangian is
\begin{equation}
    \label{eq:DMLagrangian}
\mathcal{L}=\bar{\chi}(i\slashed{D}  - m_\chi) \chi - \frac{1}{4} F'_{\mu \nu} F'^{\mu \nu}  -\frac{\epsilon}{2} F_{\mu \nu} F'^{\mu \nu} + \frac{m_{A'}}{2} A'^\mu A'_\mu ,
\end{equation}
with dark matter field $\chi$, of unit $U(1)_D$ charge, dark photon field $A'_\mu$, dark photon field strength tensor $F'_{\mu \nu}$, and ordinary photon field strength tensor $F^{\mu \nu}$. The covariant derivative $D_\mu = \partial_\mu  + i e_D A'_\mu $ has gauge coupling $e_D$, and ``dark fine structure constant"   $\alpha_D \equiv e_D^2/(4 \pi)$.

\begin{figure*}
    \begin{center}
        \includegraphics[scale=1.0]{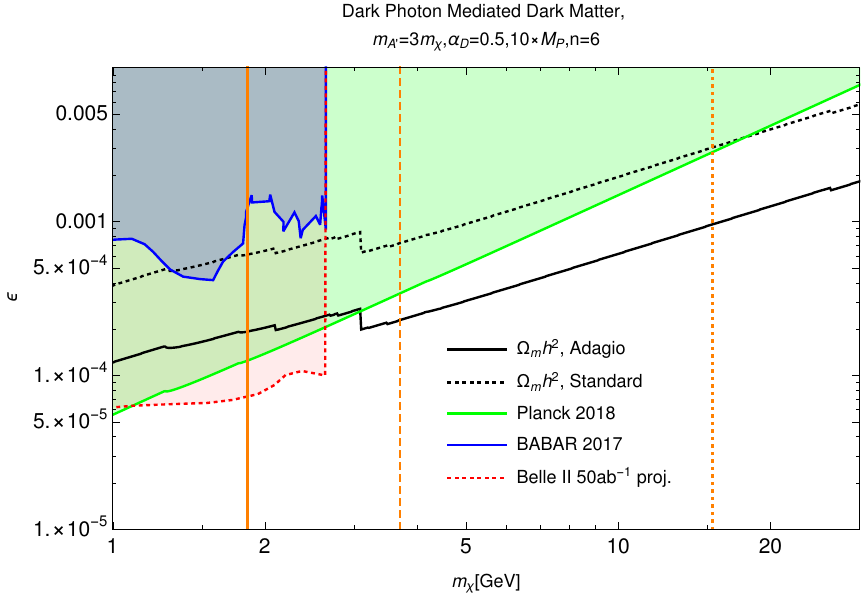}
    \end{center}
    \caption{Plot of dark matter mass $m_\chi$ versus kinetic mixing $\epsilon$ in a dark photon mediator model with $\alpha_D=0.5$ and dark photon mass $m_{A'}=3m_\chi$. Constraints from Planck \cite{Planck:2018vyg} (green), BABAR \cite{BaBar:2017tiz} mono-$\gamma$ searches (blue), and projected reach from Belle II \cite{Duerr:2019dmv} mono-$\gamma$ searches (red dotted) are shown. Curves showing the parameter space that reproduces the observed relic abundance for standard cosmology (black, dotted) and for our Adagio cosmology with 10 times larger $\mP$ (black, solid) in the early universe. The constraint on $T_{\rm min}$ is shown by the vertical orange solid line, $T_{\rm max}$ for $M_F=13$ TeV (within LHC reach) corresponds to the vertical orange dashed line, and $T_{\rm max}$ for $M_F=50$ TeV (within FCC-hh reach) is the vertical orange dotted line.}
    \label{fig:darkmatter1}
\end{figure*}

\begin{figure*}
    \begin{center}
        \includegraphics[scale=1.0]{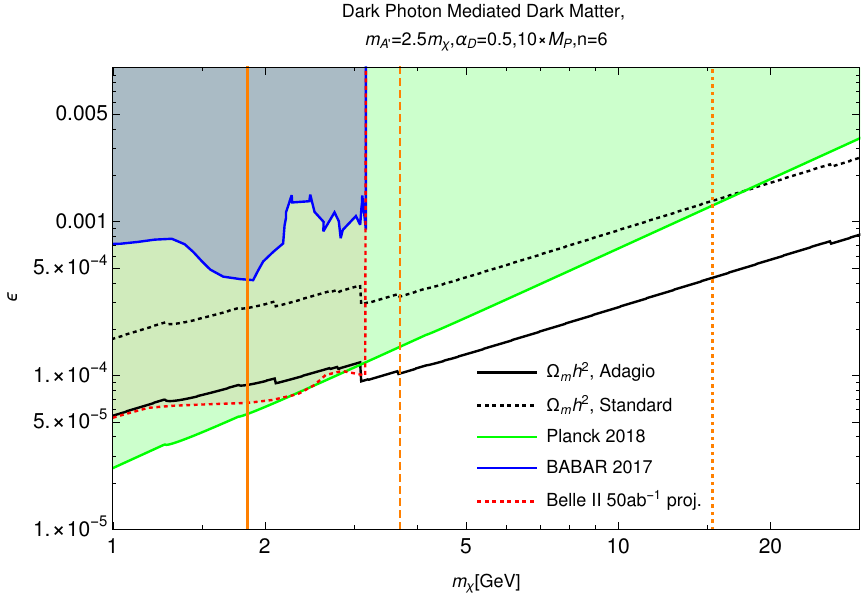}
    \end{center}
    \caption{Same as in Fig.~\ref{fig:darkmatter1} except with parameters $\alpha_D=0.5$ and dark photon mass $m_{A'}=2.5m_\chi$.}
    \label{fig:darkmatter2}
\end{figure*}

\begin{figure*}
    \begin{center}
        \includegraphics[scale=1.0]{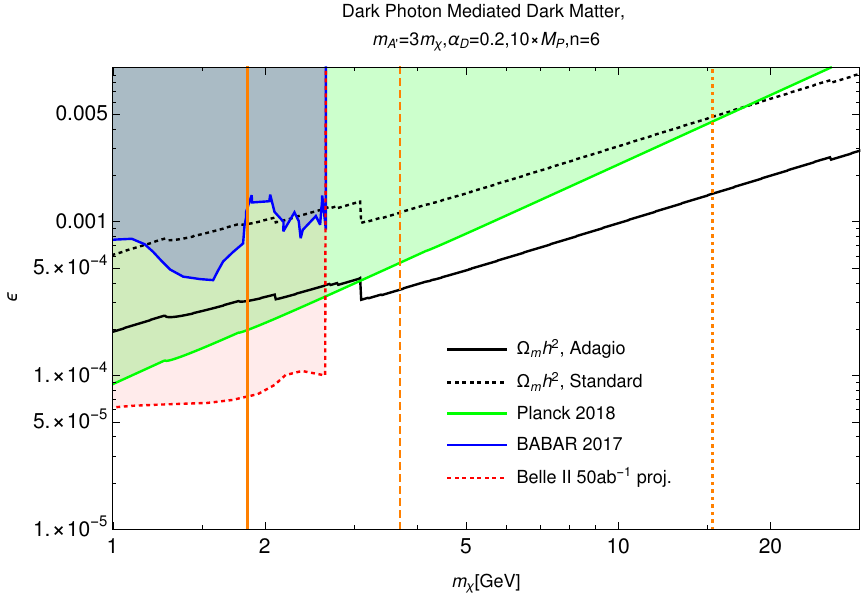}
    \end{center}
    \caption{Same as in Fig.~\ref{fig:darkmatter1} except with parameters $\alpha_D=0.2$ and dark photon mass $m_{A'}=3m_\chi$.}
    \label{fig:darkmatter3}
\end{figure*}

When $2m_\chi < m_{A'}$, the annihilation of $\chi$ is through an off-shell dark photon to SM charged particles. The $s$-wave thermal cross section for this annihilation to some charged particle with mass $m_0$, charge $Q$, and number of colors $n_c$ is given by (see {\it e.g.} Ref.~\cite{Srednicki:1988ce} for the general formalism, Ref.~\cite{Dutra:2018gmv} for the treatment applicable to dark photons, and Ref.~\cite{Fabbrichesi:2020wbt} for a simplified expression in the limit $m_0 \ll m_\chi$)
\begin{equation}
    \left<\sigma v \right> = 8 \pi \alpha \alpha_D \epsilon^2 n_c Q^2 \sqrt{1-\frac{m_{0}^2}{m_\chi^2}} \frac{2 m_\chi^2 + m_{0}^2}{(m_{A'}^2-4 m_\chi^2)^2} .
\end{equation}
In the very early universe, this annihilation cross section, together with the Hubble expansion, sets the relic abundance via thermal freeze-out, and at later times, these same annihilations will affect the CMB. On the other hand, the dark photon can be produced on-shell in collisions of SM particles at colliders, with the dark photon then decaying invisibly to dark matter almost 100\% of the time. This leads to the search channel of mono-$\gamma$ plus missing energy at $e^+e^-$ colliders. 

To determine the relic density, the following equations from Ref.~\cite{Kolb:1990vq} are used:
\begin{eqnarray}
    \label{xf}
    x_f &=& \ln\left[0.038(j+1)(g/\sqrt{g_*})\mP m_\chi \sigma_0\right]  \\
    &-& (j+1/2) \ln\left\{\ln\left[0.038(j+1)(g/\sqrt{g_*})\mP m_\chi \sigma_0\right]\right\} , \nonumber
\end{eqnarray}
where $x_f=m_\chi/T_f$ determines the freeze-out temperature $T_f$,  $g$ is the internal degrees of freedom of the relic, $g_*$ counts the relativistic degrees of freedom at freeze-out; for $s$-wave annihilation $j=0$ and $\sigma_0=\left<\sigma v \right>$.  The relic energy density in $\chi$ is then given by
\begin{equation}
\label{relicdensity}
    \Omega_\chi h^2 = 1.07 \times 10^9 \frac{(j+1) x_f^{j+1} \mathrm{GeV}^{-1}}{\sqrt{g_*} \,\mP \sigma_0}\,,
\end{equation}
where $h\approx 0.67$ \cite{ParticleDataGroup:2022pth}. For the thermal freeze-out mechanism to work, it is required that
\beq
\label{temprange}
T_{\rm{min}} < T_f < T_{\rm{max}}.
\eeq
The cold dark matter energy density of the Universe is observed to be $\Omega_{\rm CDM} h^2 = 0.12$ \cite{ParticleDataGroup:2022pth}. In our Adagio scenario, $\mP$ at the time of freeze-out is given by

\begin{equation}
\label{mplanck}
\mP = M_{\rm P,0} \kappa^{n/2},
\end{equation}
where $\kappa$ is as in \eq{kappa} and $M_{\rm P,0} \approx 1.2 \times 10^{19}$~GeV is the value of the 4-dimensional  effective Planck mass today. Since both \eq{xf} and \eq{relicdensity} depend on the  product $\mP \sigma_0$, a lower cross section can be exactly compensated by a larger Planck mass to produce the same relic abundance at the same freeze-out temperature. 

Figure~\ref{fig:darkmatter1} shows how the Adagio cosmology modifies the parameter space that leads to thermal relic dark matter, along with the relevant constraints from CMB measurements from Planck \cite{Planck:2018vyg}, mono-$\gamma$ searches from BABAR \cite{BaBar:2017tiz}, and projections for mono-$\gamma$ searches at Belle II \cite{Duerr:2019dmv} for the choice of model parameters $\alpha_D=0.5$ and $m_{A'}=3m_\chi$. Figure~\ref{fig:darkmatter2} shows the same curves for the choice of model parameters $\alpha_D=0.5$ and $m_{A'}=2.5m_\chi$. Figure~\ref{fig:darkmatter3} also shows similarly for the choice of model parameters $\alpha_D=0.2$ and $m_{A'}=3m_\chi$. Based on the allowed temperature range from \eq{temprange}, $m_\chi$ could be as large as $x_f T_{\rm{max}}$, which for typical values of $x_f \sim 20$ and $T_{\rm{max}}\sim 1$~GeV (for $M_F = 50$~TeV) leads to $m_\chi \lsim 20$~GeV.

\section{Possible Signals}

The main motivation for large extra dimensions is to solve the hierarchy problem by placing the electroweak scale close to the fundamental scale of gravity $M_F$. In this case, quantum gravity effects can be searched for at colliders. A variety of ATLAS \cite{ATLAS:2021uds,ATLAS:2023vat} and CMS \cite{CMS:2018nlk,CMS:2022fsw} searches have constrained the fundamental scale to be well above the TeV scale, with the strongest limits requiring $M_F\gsim 9.2$~TeV \cite{CMS:2018nlk}.  The ultimate LHC reach for the fundamental scale is at $M_F=13$~TeV \cite{ATLAS:2023vat}.  As center of mass energy is critical for reaching much larger values of $M_F$, we expect that a future hadron collider with center of mass energy of 100 TeV \cite{FCC:2018vvp} would be able to access $M_F\lsim 100$~TeV. 

 The large extra dimension framework naturally points to thermal relic DM around the GeV scale, which can be made compatible with the CMB constraints in the Adagio scenario.  The dark photon mediator can be searched for experimentally. While there are many collider searches for dark photons, in this scenario, the dark photon decays to dark matter instead of to SM final states. Searches for mono-$\gamma$ at Belle II \cite{Duerr:2019dmv} can, in principle, probe some of the relevant parameter space in our scenario, as seen in Figs.~\ref{fig:darkmatter1},~\ref{fig:darkmatter2},~and~\ref{fig:darkmatter3}, but that region of parameters is disfavored by the Planck limits.  
 
Most of the mass region in our dark photon mediated realization of the Adagio scenario will require higher energy, with comparable integrated luminosity to that of Belle II of 50~ab$^{-1}$. A future lepton collider with high luminosity, such as FCC-ee or CEPC, could search for dark photons beyond the mass reach of Belle II \cite{Narain:2022qud}.  We note that the cross section for $e^+ e^- \to \gamma A'$, in the limit $m_{A'}^2\ll s$, scales as $1/s$ \cite{Berends:1986yy}, where $s$ is the center of mass energy.  For a lepton collider with $s=m_Z^2$, where $m_Z=91.2$~GeV is the $Z$ vector boson mass, integrated luminosities of $\ord{100~\text{ab}^{-1}}$ have been envisioned \cite{Narain:2022qud}.  Since the cross section for $A'$ production at such a facility would be $\sim 100$ times smaller, compared to that at Belle II, we then expect a sensitivity to $\epsilon$ that is $\sim 10$ times worse.  Hence, for a future circular lepton collider operating at the $Z$ pole, we expect a sensitivity to $\epsilon \sim 5\times 10^{-4}$, for $m_{A'}\gsim 10$~GeV, corresponding to $m_\chi\gsim 3$~GeV in Fig.~\ref{fig:darkmatter1}. 

\section{Potential Alternative Utility}

Here, we will examine ``freeze-in" \cite{Hall:2009bx} as a possible alternative mechanism for the production of DM.  In this framework, DM and its associated interactions are never in thermal equilibrium.  This points to very feeble interactions between the visible  components of the cosmic energy density and the dark sector.  It is interesting that such a connection between the two sectors could  actually be motivated by astrophysical  data that seem to  favor non-standard cooling mechanisms for stellar objects \cite{Giannotti:2015kwo}.  This could be realized through the coupling of electrons, for example, to a light boson.  

Let us, for simplicity, assume that a light scalar $\phi$, in the keV regime,  couples to electrons with strength $y_e$.  One can roughly take $y_e\sim 10^{-15}$ \cite{Knapen:2017xzo} to be in the regime of interest for a possible explanation of anomalous stellar cooling hints.  A rough estimate of the freeze-in abundance $Y_\chi\equiv n_\chi/s$ produced via the light mediator $\phi$, where $n_\chi$ is the DM number density and $s\sim g_* T^3$ is the entropy density, can be obtained 
from 
\beq
Y_\chi \sim w \,y_e^2
y_\chi^2 \frac{\mP}{g_*^{3/2} m_\chi}\,,
\label{Ychi}
\eeq
with $w$ a numerical factor of $\ord{\pi^{-6}}$ \cite{Hall:2009bx} and  $y_\chi$  the coupling of $\phi$ to DM $\chi$.  For $m_\chi \sim 0.1$~GeV, as an example, DM self-interaction limits require $y_\chi \lsim 10^{-3}$ \cite{Knapen:2017xzo}.  For $g_*\sim 10$, using $m_\chi\sim 0.1$~GeV, for example, we see that $Y_\chi$ is much smaller than the $\sim 10^{-9}$ required.   

How about an interaction with muons?  Let us assume the coupling $y_\mu \phi \bar \mu \mu$.  One can estimate the rate for $\mu^+\mu^- \to \gamma \phi$ as $\sim \alpha y_\mu^2 T$.   Requiring this process to be out of equilibrium (for a freeze-in scenario and to avoid overproducing $\phi$, which could act as extra radiation and cause tension with BBN), would yield $y_\mu \lsim 10^{-9}$ and hence we may adopt $y_\mu \sim 10^{-10}$.  Using \eq{Ychi} with $y_e\to y_\mu$, we find $Y_\chi \sim 10^{-10}$, which is about $\ord{10}$ too low.  However, in an Adagio scenario with $\mP\to  \ord{10}\mP$, one may accommodate a freeze-in mechanism using muon initial states.  At the same time, the  coupling of $\phi$ to electrons may provide an explanation of anomalous stellar cooling, mentioned above.  

Here, we also note that for $m_\chi\sim 0.1$~GeV we may assume that the reheat temperature is $\sim 0.1$~GeV.  In that case, since the mass of the tau lepton $m_\tau \approx 1.8$~GeV  there would be a  suppressed thermal $\tau$ population.  Thus, one may assume that its coupling to $\phi$ is larger  than that to muons, $g_\tau\sim 10^{-8}$, without overproducing $\phi$.  A 2-loop diagram can induce \cite{Davoudiasl:2018ltz} 
\beq
\delta g_e\sim  \frac{g_\ell\,\alpha^2}{16 \pi^2} (m_e/m_\ell), 
\label{deltage}
\eeq
where $\ell = \mu, \tau$.  Here, a roughly  $m_\ell^2$ scaling of lepton  couplings to $\phi$ has been assumed, as a possibility.

For $g_\mu (g_\tau) \sim 10^{-10}  (10^{-8})$, we then find $\delta g_e\sim 10^{-18}(10^{-17})$, which does not exceed $g_e$ required by the  stellar cooling hints \cite{Knapen:2017xzo}.  Therefore, the above can  represent  phenomenologically consistent choices of parameters.

More detailed calculations  are needed for more reliable estimates and the preceding discussion is only meant to elucidate another potential  application of our Adagio cosmological scenario.

\section{Concluding Remarks}

We have shown how extra-dimensional models can realize a changing $\mP$, slowing the timescales of early universe cosmology. Using this Adagio mechanism to reduce the Hubble expansion in the early universe, GeV-scale thermal relic dark matter, disfavored by CMB constraints, becomes viable again. Naturalness arguments that the electroweak scale  should not be too separated from the fundamental scale of gravity also point to the GeV scale for thermal relics in this scenario.  In such a case, the LHC or future hadron colliders can directly look for quantum gravity effects. This new avenue for producing thermal relics with the correct abundance provides a new motivation for GeV-scale dark sector searches.

Though we focused on low mass thermal relic DM production, an Adagio scenario can, in principle, be implemented in other contexts as well.  This could potentially affect what is often called the unitarity bound on the thermal relic DM mass, which requires it to be below $\sim 100$~TeV \cite{Griest:1989wd}.  If the minimum requisite annihilation cross section can be lowered below the canonical values, one could entertain DM masses above this bound. However, this scenario would require $M_F$ to be much larger than that considered in this work to allow  reheating to much higher temperatures.  

Another potential consequence of a smaller Hubble rate, corresponding to larger causally connected volumes, could be in the relation between the temperature at which primordial black holes may form in the early Universe and their typical masses.  In the presence of an Adagio interlude, one would generally expect larger masses for such black holes, given the larger collapsing Hubble volume, at a given temperature.

Of course, one may also entertain the opposite `Allegro' scenario, where the Hubble rate is larger than the standard value as a function of temperature (or energy density, in general).  This could, for example, allow larger cross sections for thermal relics than the typical expectation, providing other alternatives for viable DM models.  However, we will not discuss this possibility further here and leave it for future work.

%%%%%%%%%%%%%%%%%%%
\begin{acknowledgments}
Work supported by the US Department of Energy under Grant Contract DE-SC0012704. Digital data files to reproduce the figures are available at \url{https://quark.phy.bnl.gov/Digital_Data_Archive/AdagioThermalRelics/}.
\end{acknowledgments}

\bibliography{Adagio-refs}

\end{document}